# Upright to supine image registration and contour propagation for thoracic patients

M.C. Martire[1,2], L. Volz[1], C. Galeone[1,2,3], M. Durante[1,4], M. Pankuch[5], C. Graeff[1,2]

[1] Biophysics, GSI Helmholtzzentrum für Schwerionenforschung, Darmstadt, Germany;

[2] Department of Electrical Engineering and Information Technology, TU Darmstadt, Germany

[3] University of Torino, Torino, Italy

[4] Institute for Condensed Matter Physics, TU Darmstadt, Germany;

[5] Northwestern Medicine Proton Center, Warrenville, Illinois, USA;

E-mail: m.c.martire@gsi.de

**Abstract**

Purpose: A renewed interest in upright therapy is currently driven by the availability of upright positioning and imaging systems. Aside from reducing facility cost, upright patient positioning possibly provides clinical advantages. The comparison between upright and supine particle therapy treatment plans can be biased through multiple variables, such as differences in the target contouring on the two CTs. We present a method for upright and supine CT registration and structures propagation, together with the investigation of an AI-based contouring tool for upright images.
Methods: Six paired 4DCTs from Proton Therapy Collaboration Group (PCG) registry were available from the Northwestern Medicine Proton Centre. Native deformable image registration (DIR) is challenged by the different patient anatomy between postures, causing artefacts in the warped images. To achieve high quality contour propagation, we propose the construction of a region of interest (ROI) covering the ribcage volume to overcome this problem. As no target contour ground truth was available, the registration quality analysis (QA) was performed on lung structures, for which dice score coefficient (DSC) and average Hausdorff distance (AHD) is reported. The TotalSegmentator tool, trained on supine dataset, was applied on upright images, verified against lung structures and used as additional comparison for contour propagation.
Results: The TotalSegmentator QA results in a maximum AHD of 2mm and a minimum DSC of 0.94. An average AHD of 1.5mm and 1.6mm, and an average DSC of 0.95 and 0.94 were obtained comparing the propagated volumes to manually contoured and AI structures, respectively. All AHD values are smaller than the CT slice distances (2.5 mm).
Conclusions: The developed DIR framework allows for accurate target propagation between upright and supine images. It defines the first step to compare upright and supine therapy of thoracic patients and enables the application of image fusion techniques in the upright therapy field.

Keywords: upright radiotherapy, image registration, thoracic patients

# 1. Introduction

The interest in Particle Therapy (PT), as an advanced form of radiotherapy, widely increased over the last decades [1]. The dose deposition of charged particles, characterised by a low entrance dose and a sharp Bragg peak (BP), allows for a precise target coverage with enhanced normal tissue sparing. With particles heavier than protons, like carbon ions, together with a narrower depth dose deposition, a smaller lateral beam spread, and range straggling and a higher biological effectiveness can be exploited [2]. Thus, carbon radiation therapy (CRT) has become a promising modality to treat radioresistant and metastatic tumors, such as non-small cell lung cancers (NSCLC) [3]. Nevertheless, physical and biological advantages of PT are penalised by higher technological hindrances, especially for heavy ions treatments. In particular, a major investment for a CRT facility is related to the massive and expensive gantry [4]. Gantry-less treatment solutions, such as upright particle therapy, make CRT accessible on a larger scale. Upright patient positioning would facilitate the rotation of the patient in front of a fixed beamline for realising different treatment angles. Without the need for a gantry, the facility cost, total footprint and total shielding material would all be reduced.

While upright patient positioning has been investigated already at the beginning of heavy ion therapy, it recently received renewed interest with the introduction of commercial solutions for patient positioning systems and upright CT imaging [5], [6]. Upright treatments for ocular and head and neck tumors are well established [7], [8] and have been in practise for decades [9]. But when the patient moves from the supine to the upright position, the action of gravity changes size, shape and position of both abdominal and chest organs. Therefore, before clinical application, the dosimetric accuracy of upright radiotherapy for these anatomical sites needs to be proven. Lung volume, rib cage position and breathing motion amplitude variation also entail anatomical changes in the abdominal region [10]. Multiple studies investigated and demonstrated changes in abdominal organs between the supine and upright body posture [10], [11], [12], [13].

Changes in patient anatomy and physiologic patterns, such as the breathing motion, directly affect the radiotherapy clinical outcome. An increased lung volume, and a decreased motion amplitude were widely investigated [10], [14], [15], [16]. A decreased motion related dose inaccuracy is a possible advantage for radiotherapy of thoracic patients and for CRT, because of its sensitivity to tissue density changes. However, the magnitude of these observations is very patient-specific and no statistically resilient study on large patient cohorts is yet available comparing the dosimetric differences between upright and supine postures. Therefore, the clinical feasibility of upright therapy to thoracic patients remains unclear.

Dosimetric comparison studies are challenged by multiple factors introducing possible bias: differences in CT quality between upright and supine scans, differences in assumed setup uncertainty, differences in breathing motion, and in particular differences in the target contour and critical structures all can have a tremendous impact on the conclusion driven from such work. In order to provide a fair comparison, the treatment plans should at least be optimized on comparable target geometries. For available data, the quality in target delineation, however, can vary between the two positions, and typically the posture finally chosen for treatment received a more detailed contouring. In addition, because of a lack of MRI and PET scans for upright positions, these imaging techniques can't be exploited to support target delineation, which is instead solely performed on diagnostic CT images. Schreuder et al. [17], investigated image synthesis techniques to overcome this problem. They present intra-modality image registration methods, to obtain upright CT and supine MRI deformed in upright geometry for volume of interest (VOI) delineation and treatment planning. Their methods require the MRI image, in addition to the CT, not always acquired for lung patients, and multiple steps to also obtain the synthetic upright MRI.

To perform a high quality upright and supine deformable image registration (DIR) and VOI propagation, it is not only important to achieve a proper contour delineation, but it is paramount to perform target propagation and consequently robust comparative dosimetric analyses between the two patient positioning. The considerable anatomic differences when the patient changes treatment position, compromise the DIR output, with a consequent poor quality target propagation.

To overcome this challenge, we developed a method for accurate DIR of paired CT images of the same patient in upright and supine, by defining a region of interest (ROI) around the rib cage and the tumor structure. Moreover, we tested the applicability of an openly available deep learning tool [18], trained on a supine CT dataset, for organ segmentation on upright CT for the first time.

**2. Materials and Methods**

In Figure 1 the method sketch is shown.

*2.1 Patient data*

Paired 4DCTs of six thoracic cancer patients, treated under the Proton Collaborative Group (PCG) registry, were available from the Northwestern Medicine Proton Centre (NMPC) in Warrenville, IL, USA. Both upright CT (600 mm FOV, 3 mm slice distance, 1 mm resolution, 10 motion phases) and supine CT (650 mm FOV, 2.5 mm slice distance, 1 mm resolution, 10 motion phases) were provided with segmented structures. Those VOIs needed for the treatment of the respective patient were manually contoured (M.C.) by the NMPC clinical staff. Specifically, the heart, lung and internal target volume (ITV) structures were available for each patient. Patients target position and volume, and lung and heart volumes are listed in Table 1 for both body positions. For two of the patients (patient P2 and P4) the target was only contoured on the supine 4DCT and the values shown in Table 1 refer to the propagated structures.

| Patient | Target position | Target volume (cc) | | Lung volume (cc) | | Heart volume (cc) | |
|---|---|---|---|---|---|---|---|
| | | Supine | Upright | Supine | Upright | Supine | Upright |
| P1 | Upper RL | 60 | 65 | 2259 | 2360 | 858 | 782 |
| P2 | Lower LL | 20 | 21* | 3921 | 4586 | 706 | 633 |
| P3 | Mediastinum | 381 | 340 | 3381 | 3957 | 497 | 590 |
| P4 | Mediastinum | 60 | 56* | 3251 | 3473 | 794 | 915 |
| P5 | Upper LL | 273 | 244 | 2643 | 3163 | 717 | 842 |
| P6 | Mediastinum | 223 | 223 | 3251 | 3109 | 695 | 722 |

Table 1. Patients characteristics in upright and supine position. Target position and target, lung and heart volumes are listed. (RL= Right Lung, LL= Left Lung)

*2.2 Deformable image registration and target propagation*

Upright to supine DIR is challenged by the considerable differences in patient anatomy that come with changing the posture. In particular, the position of the arms (arms up in supine position, arms down in upright posture), sagging of adipose tissue, shifting of internal organs and a different spine curvature presented a key registration challenge. To overcome these issues, from the CT in the reference breathing phase (end-inhale phase) for each supine and upright 4DCT set, a new CT was created defining a ROI around the rib cage and the lung volumes, surrounding the ITV. The ROI was defined to include high contrast structures, such as the rib cage and the lung, but excluding those body regions that highly differ between the upright and the supine CT of the same patient, and consequently jeopardize registration quality. On the newly created ROI the DIR was performed with Plastimatch [19] in both directions (from the supine volume to the seated one and vice versa) and the vector fields were used to propagate the ITV.

*2.3 Quality Analysis*

Registration quality analysis (QA) was performed in 3DSlicer [20] comparing ITV shape, dimension, and position in the thorax (qualitative analysis), and through the DSC and the AHD metrics (quantitative analysis) [21]. One of the challenges associated with comparative datasets is that VOI contouring quality may differ between modalities. In particular, the posture eventually chosen for treatment may be more diligently contoured compared to the other posture. As such, no ground truth was available for the tumor volume and the DSC and AHD were computed using the lung VOI. The M.C. lung structures were propagated using the previously obtained vector fields, and for both postures, the DSC and AHD values were calculated comparing the propagated VOI and the M.C. one.

In addition, the TotalSegmentator AI tool [18] was used to obtain automatically contoured (A.C.) structures, both on supine and upright CTs. The TotalSegmentator tool was used as is, i.e. using the model pre-trained on patient CT data sets where the patients were in recumbent positioning. TotalSegmentator was applied for the first time in this study on upright CTs, and as such, its performance was benchmarked using the M.C. lung structure as ground truth. For this purpose, DSC and AHD values were computed on paired integral CT, whereas the registration QA was only performed inside the ROI. The A.C lung VOI was used as additional comparison for benchmarking the contour propagation.

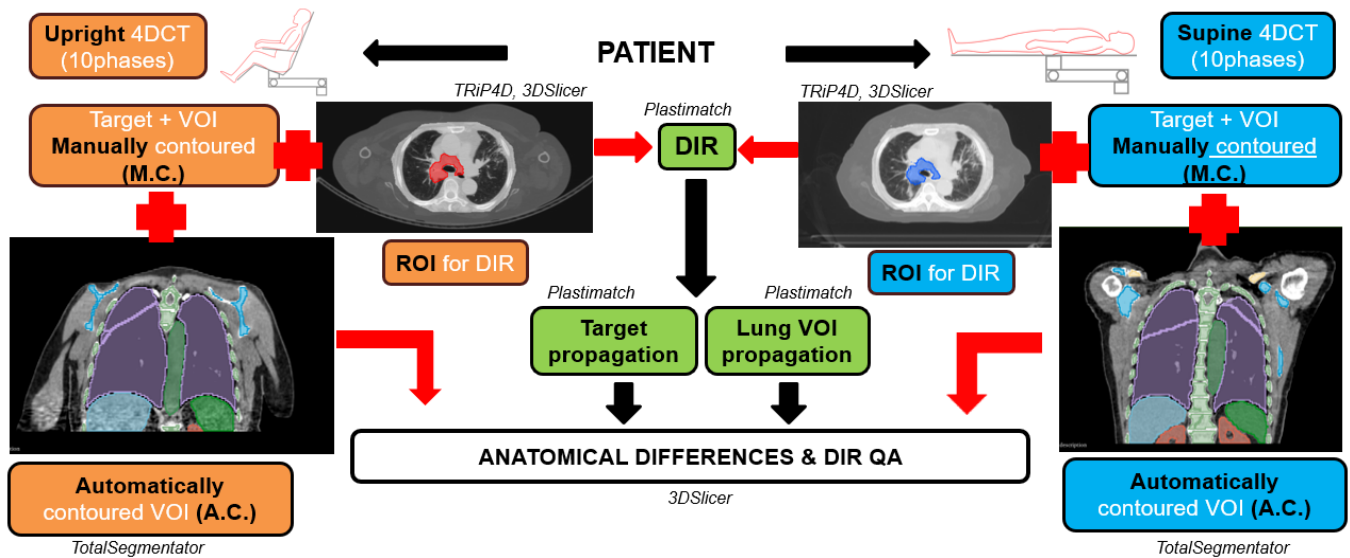

Figure 1. Registration and structure propagation method sketch

## 3. Results

*3.1 Anatomical differences and DIR*

The method presented was developed to overcome significant anatomical differences between supine and seated CTs, the main hurdle to achieve a proper DIR quality. In Figure 2, upright and supine anatomical differences are shown for patient P3. The reference phase CT is shown in coronal (2.a, 2.b), sagittal (2.c, 2.d) and axial (2.e, 3.f) view both for supine (2.a, 2.c, 2.e) and upright (2.b, 2.d, 2.f) position. Images in the first row highlight position differences, e.g., of the arms, whereas the sagittal view is used to show the shape, size and position differences between postures for the M.C. ITV (red) and heart (green) structures. Notably, the heart VOI is artificially cut at the border with the target VOI, in both volumes, same as the ITV contour at the edge of the heart on the upright CT. The images of the last row depict the ROI for DIR in the foreground, and the original CT in the background. In addition, the propagated ITV on the upright ROI is shown (Figure 2.f), matching in shape and position the M.C. ITV on the supine volume (Figure 2.e).

Moreover, composite images of Figure 3, computed in Matlab [22], show the DIR output improvement achieved with the ROI definition. Where the original and warped volumes have the same intensity, the composite image is coloured in grey, whereas different intensity regions are depicted in green and magenta. The difference between the DIR output on the entire upright and supine CTs can be qualitatively assessed (Figure 3.a). The registration quality improvement, when performed inside the ROI, becomes apparent in Figure 3.b.

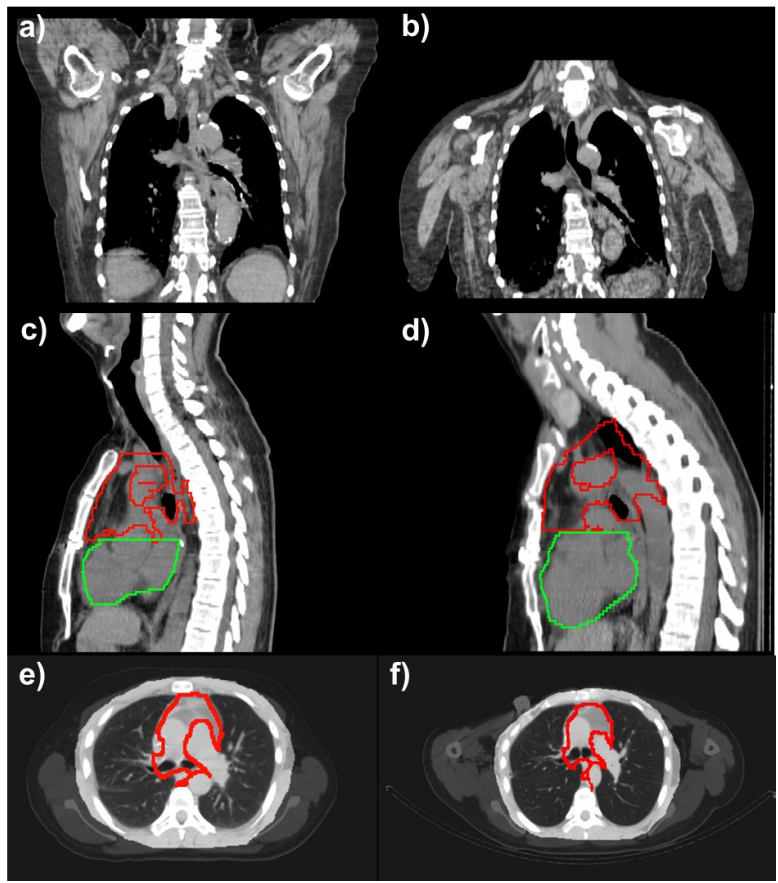

Figure 2. Patient P3 supine (a,c,e) and upright (b,d,f) reference CT in coronal (a,b), sagittal (c,d) and axial (e,f) view. Body position (a,b) and target (red) and heart (green) contour (c,d) differences are highlited. The ROI and the original (e) and propagated (f) target are also shown.

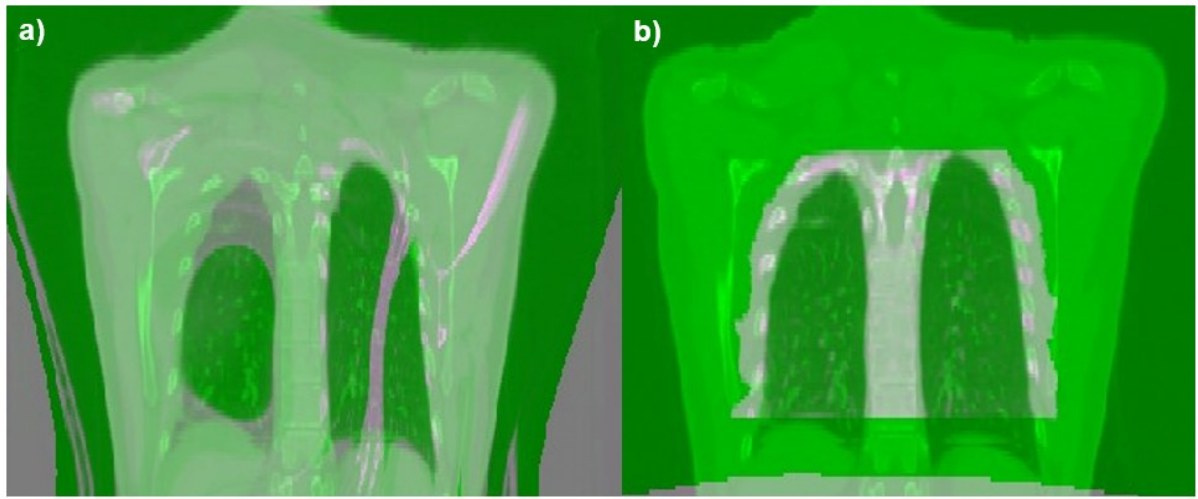

Figure 3. Composite images of the original and warped volumes, computed on the entire upright and supine CT (a), and confined inside the ROI (b). Grey areas define a similar image intensity, while different intensities are depicted in green and magenta.

*3.2 TotalSegmentator tool*

The TotalSegmentator tool was applied to all the dataset CTs and, along with the output on the supine volume (Figure 4.a, 4.b), the result for the upright CT (Figure 4.c, 4.d) is shown on the reference volume for patient P2. Although it is not possible to compare matching slices, both because of a different total number of slices in the upright and supine CT, and because of the morphological differences, Figure 4 allows to qualitatively assess the deep-learning tool output quality for both patient positions. In addition, in Table 1 of supplementary material, DSC and AHD values are shown for each patient and both for supine and upright CT. In this case, as described in section 2.3, the M.C. structures were used as reference, and the metrics were calculated on the lung VOI. All DSC are above 0.90, considered as excellent, and all the AHD values are smaller than the distance between CT slices, namely 2.5 mm and 3 mm for supine and upright CT respectively. The QA was performed both on the supine and the upright CT. For the supine cases, the average values are 0.95 and 1.60 mm for DSC and AHD respectively, whereas the QA results on the upright CT show an average 0.96 and 1.3 mm for DSC and AHD respectively. These results prove the A.C. and M.C. lung VOIs have comparable quality, allowing the use of both as ground truth for the registration and propagation QA.

*3.3 Registration and propagation QA*

In the lack of a target VOI ground truth, the QA for the DIR and contour propagation was performed on the M.C. and A.C. lung structures. For each patient, the M.C. lung VOIs were propagated in both directions, exclusively inside the ROI, and the DIR and propagation QA results are listed in Table 2 of the supplementary material. The first two columns of metrics refer to the M.C. structures used as reference, whereas the last two columns show DSC and AHD values for the case in which the A.C. VOIs are used as reference. The DSC is always above 0.90, and the AHD values are smaller than the CT slice distance. Propagated lung VOIs compared to the M.C. ones result in an average DSC and AHD values of 0.95 and 1.5 mm respectively, both considered excellent. When these metrics were calculated comparing the propagated lung structure and the AI tool output an average, among the six patients of 0.94 and 1.6mm were obtained for DSC and AHD values, respectively.

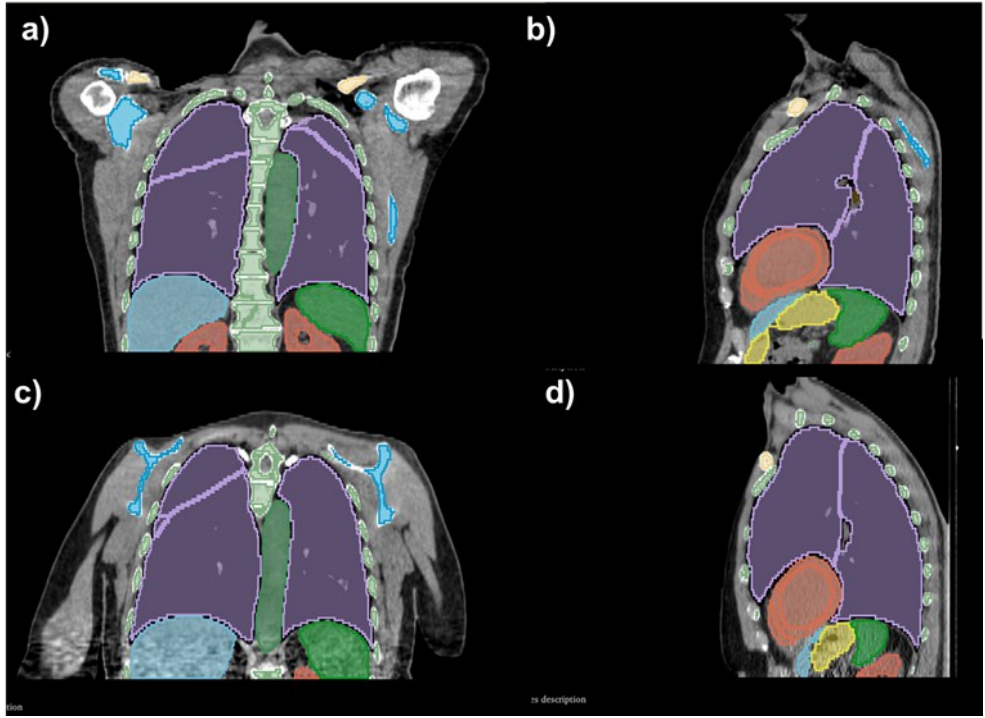

Figure 4. Patient P2 TotalSegmentator output on supine (a,b) and upright (c,d) reference CTs, in coronal (a,c) and sagittal (b,d) view.

*3.4 Propagated target volumes comparison*

As mentioned in section 2.3, the target propagation quality can only be assessed comparing structures shape, position and volume. Upright and supine M.C. ITV volumes and propagated ITV volumes, in both propagation direction, are listed in Table 2. In the first two columns, the manually segmented supine ITV is compared with the propagated one on the paired upright CT, while the opposite propagation direction is summarised in the last two columns. An average volume difference of 6% is obtained for all the six patients and both propagation directions, with a maximum of about 12% difference for upright ITV of patient P6 propagated on the supine volume, and an exact equal volume between the upright ITV of patient P5 propagated on the supine CT. In addition, in Figure 1 of the supplementary material, target structures for patient P1 (Figure 1.a and 1.b) and P6 (Figure 1.c and 1.d) are qualitatively compared. Original and propagated structures are shown in 3D view and displayed together to assess geometry and shape similarities. Patients P1 images show the propagation of the ITV originally contoured on the upright CT (white) and the propagation result on the supine volume (green), whereas the supine ITV M.C. on the reference volume (white) is displayed alongside the propagated result on the upright CT (green) for patient P6.

| Patient | Target volume (cc) (Supine→Upright) | | Target volume (cc) (Upright→Supine) | |
| --- | --- | --- | --- | --- |
| | Supine | Upright | Upright | Supine |
| P1 | 60 | 58 | 65 | 68 |
| P2 | 20 | 21 | - | - |
| P3 | 381 | 364 | 340 | 361 |
| P4 | 60 | 56 | - | - |
| P5 | 273 | 262 | 244 | 244 |
| P6 | 223 | 249 | 223 | 195 |

Table 2. M.C. and propagated target volumes in both patient positions and for both propagation directions.

## 4. Discussion

We developed a method to perform image registration between supine and upright CTs and applied it on six thoracic cancer patients. Image registration and target propagation are preliminary steps to perform resilient dosimetric comparisons between the two treatment modalities. High quality DIR is possible thanks to a rib-cage ROI defined to overcome anatomical differences between the two patient positions. Vector fields are benchmarked both using manually contoured and AI structures obtained with the deep-learning tool TotalSegmentator, tested for the first time on an upright data set.

Marano et al. [10] investigated, using the same dataset presented in this work, the thoracic anatomical changes between the upright and the supine body posture, assessing heart and lung differences of interest for therapy purposes. Structure propagation is heavily affected by these anatomical changes if thoracic CTs of the same patient, acquired in two different body positions, have to be registered. The construction of a ROI, extended to a number of slices surrounding the target volume, and limited to the rib cage area in each slice, allows to overcome these differences. For treatment planning study purposes, the target is the structure of interest that has to be propagated, however a ground truth is missing. The manually contoured target can be affected by radiation oncologist errors, and like most other extra-cranial structures, the target changes size, shape and position when the patient moves from the recumbent to the seated position. In order to understand observed differences between supine and upright positioning in future particle therapy treatment planning studies, effects from various sources need to be isolated. For example, if the geometry of the tumour differs between postures (either due to contouring differences, or differences in the actual structure), it may change the dose to the surrounding critical organs and overshadow the effects from increased lung volume or reduced breathing motion. Although, ultimately, the treatment quality is what matters, upright radiotherapy comparative clinical trials are not yet available, and resilient data will remain outstanding for some time. In the meantime, contour propagation methods, as presented in this work, will enable to better understand the various effects related to the change in posture and draw conclusions on the suitability of upright positioning for different patients.

Precise target and organs at risk contouring on upright volumes, represents the first pitfall for treatment planning purposes. Because of the lack of clinical available MRI and PET upright imaging systems for radiotherapy, the VOI definition can only rely on diagnostic CT scans, characterised by a low soft tissue contrast. Schreuder et al [17] investigated two methods, based on synthetic CT or MRI images, to overcome this problem. Multiple steps are needed, to generate a synthetic image and perform intra-modality registration. In our case, the only required step preliminar to the registration, is the definition of a ROI in both CTs. However, also Schreuder et al. suggested in their work the use of a restricted region as a possible approach to mitigate upright and supine anatomical differences.

A limitation of this work was that the QA was only based on the lung VOI, which is relatively straightforward to segment (manually and AI based) and register because of the high density contrast with the surrounding tissues. Since the vector fields are only constructed inside the ROI, the DIR QA was also restricted to this volume. Even if other structures, such as the heart, the spine and the rib cage were present in the selected ROI, only the heart was available in the original dataset for each patient, and depending on the tumor position in the lung, this could be placed outside the ROI. Moreover, the TotalSegmentator tool outputs individual segmentation for each heart sector, resulting in a considerably different VOI compared to the M.C. heart, even in supine datasets, where the heart was available only as a whole. Hence, the AI tool QA results would not be relevant for this specific VOI. This left only the lung structures for performing the DIR and A.C QA. Nevertheless, the high similarity between the propagated, M.C and A.C. contoured lung VOIs, indicated by the excellent DSC and AHD scores, gives confidence in the quality of the proposed DIR framework.

## 5. Conclusions

A DIR method for propagating the target contour and close OARs between supine and upright CT datasets was presented in this work. The developed method restricts the DIR to a ROI enclosing the ribcage of the patient, which enables it to overcome the challenges from the strong differences in patient anatomy between upright and supine positioning. The high-quality DIR between upright and supine images registration was confirmed by comparing the propagated lung VOI to AI and manually contoured lung structures. This work's most direct application will be in future radiotherapy and particle therapy dosimetric studies, aiming to compare upright and supine therapy. This method represents a significant step towards optimizing paired treatment plans and assess the clinical impact of different patient positions. In addition, the possibility to register upright and supine images makes upright image fusion techniques possible, currently only available for supine medical images. Indeed, functional imaging techniques (PET/MRI) are still not available for upright radiotherapy clinical practice and the DIR is a mandatory step for multimodality imaging workflows.

**Acknowledgements**

This project has received fundings from the PARTITUR project funded by the German Federal Ministry of education and research (02NUK076A).

# Supplementary material

| Patient | TotalSegmentator QA (Supine CT) | | TotalSegmentator QA (Upright CT) | |
|---|---|---|---|---|
| | DSC | AHD (mm) | DSC | AHD (mm) |
| P1 | 0.95 | 1.80 | 0.94 | 1.7 |
| P2 | 0.95 | 1.60 | 0.96 | 1.30 |
| P3 | 0.94 | 1.56 | 0.96 | 1.20 |
| P4 | 0.97 | 0.90 | 1 | 0.1 |
| P5 | 0.93 | 2.1 | 0.95 | 1.7 |
| P6 | 0.95 | 1.60 | 0.94 | 2 |

Table 1. TotalSegmentator QA results. For each patient, the DSC and AHD, from the M.C. and A.C. VOIs comparison are listed. The QA was performed both on the supine and on the upright reference CT.

| Propagation direction | Patient | DIR & propagation QA | | | |
|---|---|---|---|---|---|
| | | M.C. VOI vs propagated VOI | | A.C. VOI vs propagated VOI | |
| | | DSC | AHD (mm) | DSC | AHD (mm) |
| Upright→Supine | P1 | 0.93 | 1.50 | 0.92 | 1.75 |
| | P2 | 0.96 | 1.52 | 0.94 | 1.60 |
| | P3 | 0.96 | 1.36 | 0.95 | 1.75 |
| | P4 | 0.95 | 1.42 | 0.97 | 0.84 |
| | P5 | 0.92 | 2 | 0.94 | 1.80 |
| | P6 | 0.96 | 1.24 | 0.94 | 1.60 |
| Supine→Upright | P1 | 0.95 | 1.45 | 0.94 | 1.58 |
| | P2 | 0.93 | 2 | 0.93 | 1.8 |
| | P3 | 0.95 | 1.48 | 0.95 | 1.54 |
| | P4 | 0.96 | 1.46 | 0.96 | 1.59 |
| | P5 | 0.94 | 1.50 | 0.93 | 1.70 |
| | P6 | 0.96 | 1.2 | 0.94 | 1.60 |

Table 2. DIR and lung VOI propagation QA. For each patient and each propagation direction the DSC and AHD values are listed. The analysis is performed using both the M.C. and the A.C. lung VOIs are ground truth.

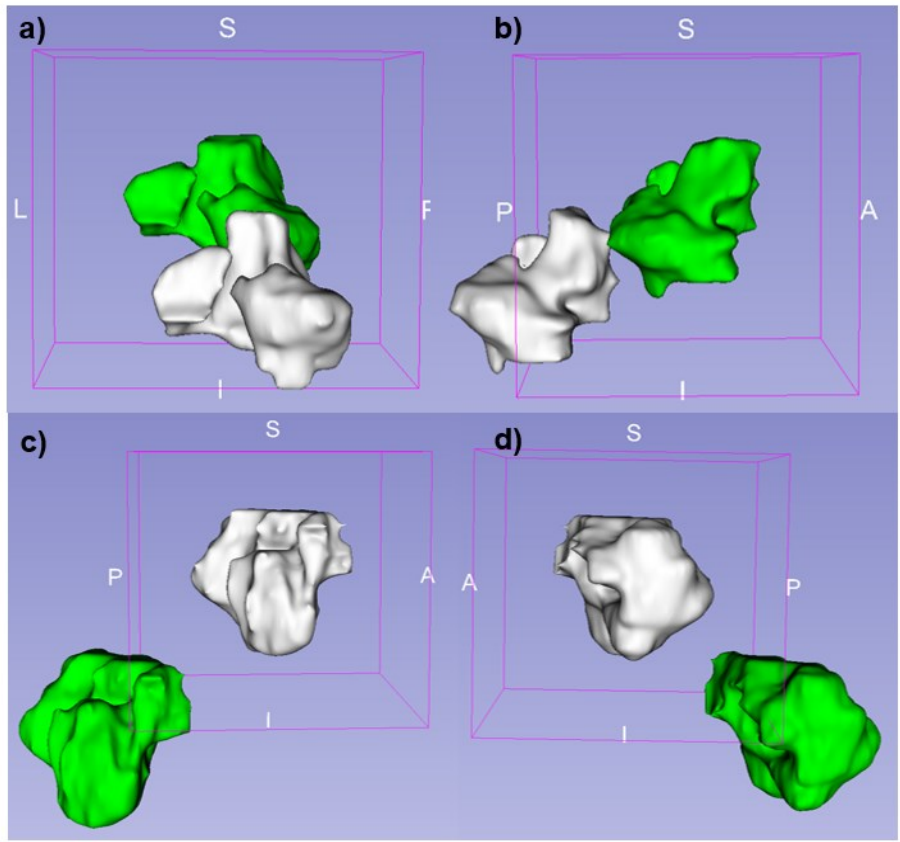

Figure 1. Patient P1 (a,b) and P6 (c,d) original (white) and propagated (green) target in 3D view. For P1, the M.C. upright target and the propagated one on the supine geometry are shown, while for P6, the M.C. supine target and the propgated one on the upright CT are shown.